\documentclass[aps,prc,twocolumn,showpacs,floats,byrevtex,superscriptaddress,nofootinbib]{revtex4}

\usepackage{epsfig}

\newcommand{\nuc}[2]{$^{#1}${#2}}
\def\etal{\emph{et al.}}
\def\be{\begin{equation}}
\def\ee{\end{equation}}
\def\lb{\langle}
\def\rb{\rangle}
\def\olp2{\lb q_i | q_{i+1}\rb}
\def\olp3{\lb q_i | q_{i+2}\rb}
\def\pb208{$^{208}$Pb}
\def\sn120{$^{120}$Sn}
\def\sm156{$^{156}$Sm}

\def\lb{\langle}
\def\rb{\rangle}
\def\bRb{\lb q| R_\theta| q'\rb}

\begin{document}

\title{Correlation energies by the generator coordinate method:\\
computational aspects for quadrupolar deformations}

\author{M. Bender}
\affiliation{Service de Physique Nucl{\'e}aire Th{\'e}orique,
             Universit{\'e} Libre de Bruxelles,
             CP 229, B-1050 Brussels, Belgium}

\author{G. F. Bertsch}
\affiliation{Department of Physics and Institute for Nuclear Theory
Box 351560, University of Washington, Seattle, WA 98195}

\author{P.-H. Heenen}
\affiliation{Service de Physique Nucl{\'e}aire Th{\'e}orique,
             Universit{\'e} Libre de Bruxelles,
             CP 229, B-1050 Brussels, Belgium}

\date{November 7 2003}

\begin{abstract}
We investigate truncation schemes to reduce the computational
cost of calculating correlations by the generator coordinate method
based on mean-field wave functions. 
As our test nuclei, we take examples for which accurate calculations
are available.  These include a strongly deformed nucleus,
$^{156}$Sm, a nucleus with strong pairing, $^{120}$Sn, the krypton
isotope chain which contains examples of soft deformations, and 
the lead isotope chain which includes the doubly magic \pb208.
We find that the Gaussian overlap approximation
for angular momentum projection is effective and reduces the 
computational cost by an order of magnitude. Cost savings in
the deformation degrees of freedom are harder to realize.  
A straightforward Gaussian overlap approximation can be applied rather
reliably to angular-momentum projected states based on configuration sets 
having the same sign deformation (prolate
or oblate), but matrix elements between prolate and oblate
deformations must be treated with more care.  We propose a two-dimensional 
GOA using a triangulation procedure to treat the general case with
both kinds of deformation.  With the computational gains from 
these approximations, it should be feasible to carry out
a systematic calculation of correlation energies for
the nuclear mass table.
\end{abstract}

\pacs{21.60.Jz, 21.10.Dr}

\maketitle

\section{Introduction}
Much progress has been made in developing a fully microscopic method to
determine nuclear binding energies. The most successful approach
in this framework \cite{go02} starts from an energy functional of the
nuclear orbitals \cite{BHR03}, which is minimized by solving mean-field
equations for the orbitals.  However, to achieve a level of accuracy 
around 700 keV on the more than 2000 known masses, some correlation 
energies beyond what can be 
subsumed within the mean-field functionals have to be introduced. 
Unfortunately, the well-established microscopic methods which could
be applied are far too costly to be used for such a large number of
nuclei.  As a consequence, in the systematics of binding energies, the
effects of correlations are estimated with phenomenological methods
whose range of validity is not apparent. Our  
goal is find microscopically founded yet easy-to-implement methods
allowing an accurate calculation of these correlations.

There are several systematic approaches to correlation energies.  Among
them,  the
generator coordinate method \cite{BHR03} is especially promising, mainly 
because of two attractive features: it is a fully variational method
and it is not limited to small-amplitude collective motion.
It has been used successfully by the Paris-Brussels collaboration for
calculating spectroscopic properties of nuclei (for representative
applications, see Refs.\ \cite{bo91,HBD93,MBW95,HVB01,va00,be03,du03,BFH03}), 
and we believe it has promise for a global microscopic theory to
systematically calculate nuclear masses.  

The most important fluctuations beyond the mean field are those
associated with pairing fluctuations and with quadrupolar deformations.
We only discuss the latter in this work; the pairing fluctuations are 
treated by number projection as in the above cited GCM studies.
At the mean-field level,
quadrupolar deformations are introduced by using a quadrupolar 
constraining field to generate deformed configurations.
To go beyond the mean field,
one mixes configurations to obtain an additional correlation energy.
This can happen in two ways. 
First, whenever the mean-field minimum is deformed, it can be interpreted as
the intrinsic state of a rotational band. The lowest mean-field
energy then corresponds to a weighted average over the members of the  
band, and one gains correlation energy by projecting onto the
ground state of the band.
Second, fluctuations of the quadrupole deformation about the
value of the mean-field energy minimum also contribute to the
correlation energy. In fact, when that minimum is spherical, they are
the only one present. Both kind of
correlations can be introduced simultaneously by
projecting good angular momentum from the intrinsic states and by
mixing configurations around the mean-field
energy minimum with the generator coordinate method
with the quadrupolar deformation as generator coordinate.
Note that because the resulting wave function
has amplitudes for deformed as well as spherical configurations,
the distinction between these two correlations becomes blurred.

\begin{figure*}[t]
\centerline{\epsfig{figure=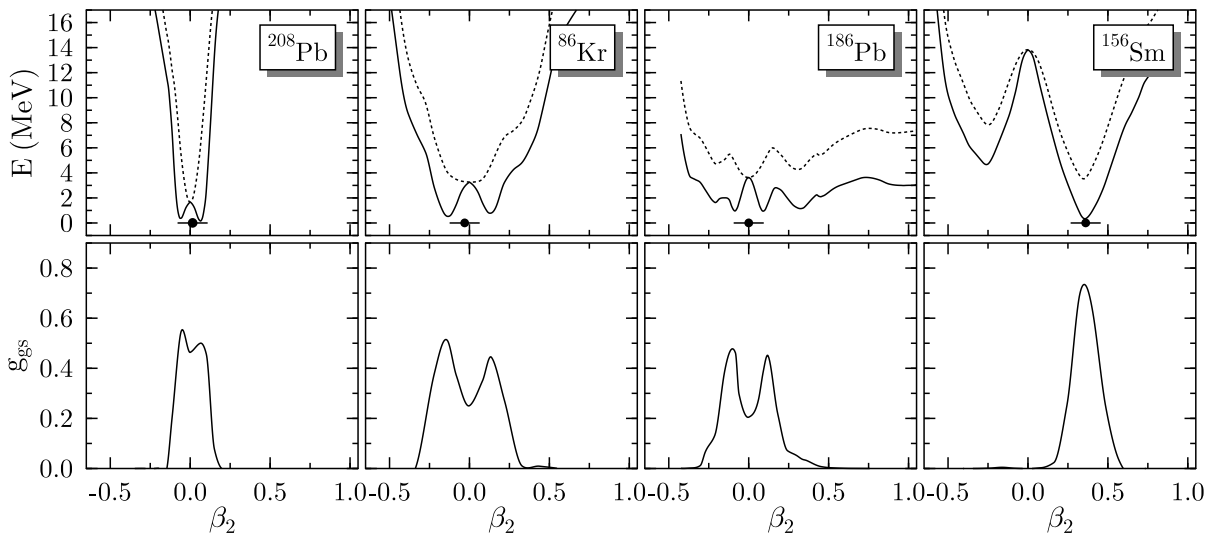}}
\caption{\label{figure1}
Overview of potential energy landscapes for selected nuclei.
See text for details.
}
\end{figure*}

In previous works \cite{du03,be03,BFH03,va00}, the procedure to introduce
these correlations consists of three steps. First one constructs
mean-field wave functions for a set of configurations that
are defined by a constraining quadrupolar field.  These configurations
may be labeled by the Bohr-Mottelson variables of two deformation
parameters $\beta$ and $\gamma$. 
From them, additional states are generated by
rotations with Euler angles $\Omega= (\phi,\theta,\chi)$ from which
are projected states of good angular momentum. The projected
states corresponding to different values of $\beta$ and $\gamma$ 
are then mixed by the GCM, which means numerically solving the Hill-Wheeler
equations in the discrete basis of the projected configuration set.
The computational issue in such calculations is in
the selection of the Euler angles and the deformation parameters
that are necessary to achieve a sufficient accuracy. Up to now,
a fairly large number of active angles ($\sim12$) were used for each
deformation.  We will show that this can be drastically simplified
by using a suitable Gaussian overlap approximation, particularly
the topological Gaussian overlap approximation advocated
in ref.\ \cite{ha02}.

Figure \ref{figure1} gives an overview of the energetics associated
with the $\beta$ degree of freedom in selected nuclei.  We have singled out 
\nuc{208}{Pb} as an example of a doubly magic nucleus, \nuc{86}{Kr}
as a soft spherical vibrator, \nuc{186}{Pb} as a soft nucleus with several
coexisting near-degenerate minima, and \nuc{156}{Sm} as a well-deformed
nucleus. The upper panel shows the energy surface, measuring the
energy with respect to that of the GCM ground state.  The dashed curve shows
the energies of the states obtained by $N$ and $Z$ number projection
from the HF+BCS configurations.  The configurations are labeled by
their intrinsic quadrupole moments\footnote{The negative values of
$\beta$ occur for oblate axially symmetric deformations; in
the Bohr-Mottelson parameterization they correspond to $\beta=|\beta_2|$
and $\gamma = 60^\circ$, 180$^\circ$, or 300$^\circ$, depending on the
choice of symmetry axis.
}
\be
\beta_2 = \frac{4 \pi}{3 A r_0^2}  
\langle r^2 \, Y_{20} \rangle
\ee
with $r_0 = 1.2 \, A^{1/3}$  fm. In the figure, the states are
connected by the line.  One
sees that the ground state is spherical except for \sm156, which 
has a strong prolate deformation.  The solid curves show the results
of angular momentum projection onto $J=0$, labeling the states by
the deformation of the state from which they were projected.
One sees that there is a net gain of energy using a deformed 
intrinsic configuration for all the nuclei including \pb208.  The final
energy, obtained by the Hill-Wheeler equations using the $Z,N,J$-projected
configurations, is at zero energy on the graph.  The dot indicates
the average deformation of the mean-field states within the 
Hill-Wheeler ground state.
The lower panel shows the relative amplitudes of the different 
configurations in the ground-state wave function.  For \sm156, the
function has a single peak and can easily be interpreted as the
zero-point motion of the $\beta$-vibration.  For the other nuclei, 
there is a strong mixing of oblate and prolate configurations.  As
we will see below, this makes a simplified treatment for spherical nuclei
somewhat more complicated than for the fully deformed nuclei.

The GOA has often been used in the past to derive the ingredients of a 
collective Bohr Hamiltonian in the spirit of the work of Kumar and 
Baranger~\cite{Kum68}. It is supplemented by several other 
approximations~\cite{BHR03,GG79}. The resulting Hamiltonian 
is five-dimensional, three dimensions corresponding to rotation and two
to quadrupolar vibrations. This approach has been followed in ref. [14]
and [15] to improve on the mean field theory of the Skyrme and Gogny 
interactions, respectively. The Lublin group has 
derived an alternative scheme starting from a Nilsson 
Hamiltonian~\cite{go85,Pil93,pr99}, which allows one to introduce several 
collective degrees of freedom. In the present work, the GOA is a
numerical approximation, in principle under control, used to evaluate 
matrix elements of the norm and Hamiltonian kernels of the generator 
coordinate method, including projection on good angular momentum. 

Most previous calculations with the angular-momentum projected
GCM have assumed axial symmetry, i.e.\ taking only configurations 
with $\gamma$ a multiple of $\pi/6$, 
and we will make the same assumption here. In practive, the
optimal spacing of configurations as a function of $\beta$ remains
to be determined, and also 
which off-diagonal matrix elements need to be calculated explicitly 
rather than estimated by the GOA. The
first question was studied in ref.\ \cite{bo90} for unprojected
matrix elements of the nucleus $^{194}$Hg.  We will see that the same
considerations apply to the projected matrix elements, provided that
the special characteristics of the 
$\beta=0$ singular point are taken into account.

In the next section, we examine several methods for
the approximate angular projection.  Our target is to design a
procedure leading to an accuracy of
a few hundred keV, the level required to improve the best present 
mean-field based approach to binding energies.  We find
that a two-point approximation gives acceptable accuracy in
all but rare circumstances.  In Section IV we discuss the
mesh of GCM configurations entering the Hill-Wheeler equation, 
and we propose a prescription for 
simplifying the space without losing the desired accuracy.

Let us now briefly describe our procedure for dealing with
pairing fluctuations.
The mean-field calculations are performed in the Hartree-Fock+BCS 
approximation using the Skyrme interaction for the
particle-hole channel field and a separate zero-range interaction
for the pairing channel. The mean-field interaction is taken to
be the SLy6 parameterization in the present work, the same as was 
used in ref.\ \cite{du03}. The pairing interaction is taken as a 
density-dependent delta function as in ref.\ \cite{ri99}; it is used with
5 MeV cut-off of the orbital space above and below the Fermi energy.
We use the same strength of the pairing as in ref.\ \cite{ri99},
(-1250 MeV fm$^3$), except for the  Kr isotopes, where it taken as
-1000 MeV fm$^3$. To avoid the phase transition of the BCS theory 
at nuclei with low densities
of state at the Fermi surface, the Lipkin-Nogami prescription is used
to generate the BCS wave functions, which are then projected 
onto fixed numbers of protons and neutrons. The last step is 
essential to have nuclide-specific predictions. We always use 
number-projected wave functions in this work, and will refer to them
as the \emph{intrinsic configurations} of the GCM to distinguish them
from the BCS configurations that do not have definite particle numbers.

\section{Angular projection}
 
The angular projection of a deformed wave function is carried
out by calculating the integral
\be
\label{exactintegral}
I_0(q,q') 
= 2 \int_{0}^1 d\cos (\theta) \bRb
\ee
and similar expressions for the matrix elements of the Hamiltonian
and other operators. 
Here we have used the axial symmetry and the symmetry with
respect to $180^\circ$ rotations, $\bRb = \lb q| R_{\pi-\theta}| q'\rb$,
to reduce the integration over Euler angles to the above single
integral. These properties can be used if the left and right wave functions
have the same symmetry axes.  We shall define our intrinsic configurations
accordingly, taking the symmetry axis along $z$.  This
requires that the oblate wave functions correspond to 
$\gamma=180^\circ$ instead of the usual $60^\circ$. 

In ref.\ \cite{du03} the integration was carried out on a 
Gauss-Legendre mesh with enough points to thoroughly sample the integrand
in the range that it is nonzero. 
This is done in such a way that using the estimate for the overlap
given in Ref.\ \cite{ba86}, the total number of abscissas in the interval
$[-1,1]$ is chosen in such a way that at least 12 abscissas can be expected
to give overlaps that are larger than $10^{-8}$ times the overlap at 
$\theta = 0$. The calculation of the integrand is started at $\theta = 0$
and stopped when the calculated overlap has indeed fallen below this
value. In this way, the choice of the Gauss-Legendre mesh is self-adjusting 
to the structure of the integrand, which scales with the dispersion of
the angular momentum and is therefore nearly constant for states with small
deformation and sharply peaked at $\theta = 0$ for well-deformed ones.

As discussed in the Appendix, the rotation operation is a rather 
costly computational task.  The Gaussian overlap approximation 
reduces the number of points to one (besides the identity), making 
it attractive for global applications.  There are 
a number of ways that the approximation can be applied.  We
shall write a generalized GOA for the overlap in the form
\be
\bRb = \lb q | q' \rb  \, e^{-c( q, q') \, F(\theta)}
\ee
where $F$ has a fixed functional dependence on $\theta$.
Since there is only parameter in 
the expression, namely $c(q,q')$, only one rotation and many-particle overlap 
has  to be carried out to determine it.  The
integral over the Gaussian is then a trivial computational
task.

The naive GOA takes $F(\theta) = \theta^2$, which was 
proposed in early papers on the subject \cite{vi71,is79}.  
Another form suggested in ref.\ \cite{ba84} is  $F= 1-\cos(\theta)$. 
We will call this the improved GOA.  It is 
designed for cranking wave functions
which break the time-reversal invariance. 
In our case here, the wave functions are invariant under time reversal
and have additional rotational symmetries as well.
With quadrupolar amplitudes to be
rotated, a form having the correct limit for small deformations
is \cite{is79}
$$
\bRb 
= \exp \big\{ - \sum c_{m,m'} \big[ \delta_{m,m'} - {\cal D}^2_{m,m'}(\Omega)
       \big] \big\}
$$
where ${\cal D}^2$ is Wigner's rotation matrix for $J=2$.
Hagino \etal\ \cite{ha02} have considered this form for the case of axial 
symmetry, taking $F$ as
\be
F = \sin^2 (\theta) \sim 1-{\cal D}^2_{0,0} .
\ee  
Testing the approximation in a three-level Lipkin model, they found an 
important gain in accuracy using the $\sin^2 (\theta)$ dependence,
which they call ``topological GOA'' (topGOA).
They also tested the approximation on the Interacting Boson Model
and found similar improvement \cite{ha03}.

\begin{figure}[t]
\centerline{\epsfig{figure=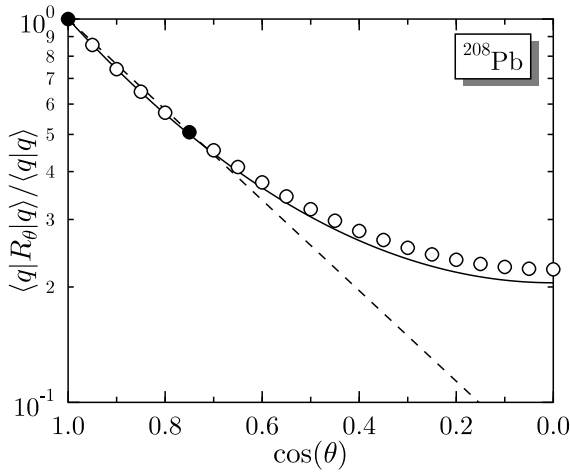}}
\caption{\label{figure2}
The overlap $\lb q | R_\theta| q \rb$ for the nucleus $^{208}$Pb
at $q=500$ fm$^2$.  The overlaps of the particle-number projected BCS intrinsic
states have been renormalized to one at $\theta=0$. The circles show the 
full calculation with the
rotated many-particle wave functions. The fit by the topGOA is
shown by the solid line and  the fit by the improved GOA is shown as
the dashed line. The $\theta=0$ point and the point shown as
the filled circle were used to make the fit.
}
\end{figure}

We show in Fig.\ \ref{figure2} the overlap $\lb q | R_\theta| q \rb$
as a function of angle for a case of moderate deformation, $^{208}$Pb 
at a deformation of -5 b ($\beta_2=-0.063$). The points were calculated
with the full many-particle wave functions.  The GOA is constructed using
the overlap at $\cos (\theta) = 0.75$, where it is close to 0.5. 
The curves show the fits with the improved GOA and with
the topGOA. While the improved GOA loses accuracy away from the
peak,  the topGOA gives a reasonable fit throughout. The projected
integral $I_0$ is off by 18$\%$ for the improved GOA and only 1.3$\%$
for the topGOA.

\begin{figure}[t]
\centerline{\epsfig{figure=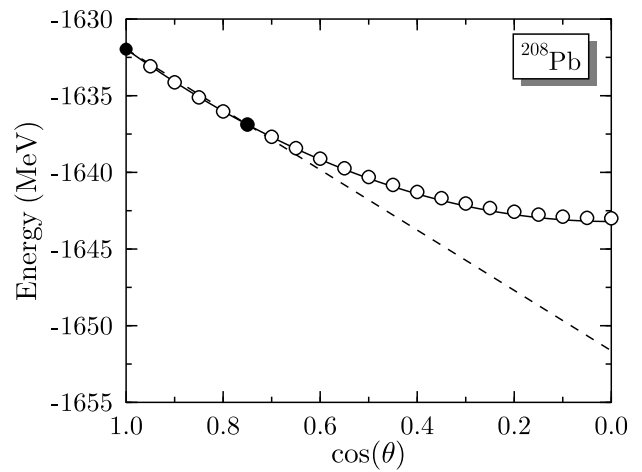}}
\caption{\label{figure3}
Hamiltonian matrix element as a function of rotation angle for 
\pb208 at $q=-500$ fm$^2$. The
circles show the calculated values with the full many-particle wave
function, and the lines are fits to the forms $h_0-h_2 \sin^2 (\theta)$ 
(solid) and $h_0 -h_2 (1.0-\cos\theta)$ (dashed). The two filled 
circles show the matrix elements that were used for the fit.
}
\end{figure}

Next we consider the Hamiltonian matrix elements and their projection.
Part of the GOA scheme as is usually
applied assumes that the Hamiltonian matrix elements can be written
as a product of the overlap times a factor that is quadratic in
the difference of generator coordinates.  For all versions of the
GOA this may be expressed
\be
\label{hgoa}
\lb q | H R_\theta | q'\rb 
= \lb q | q' \rb\,e^{-c(q,q') F(\theta)} \Big[h_0(q,q') -h_2(q,q') 
F(\theta)\Big]
\ee
The values of $h_0$ and $h_2$ may be determined from the two Hamiltonian
matrix elements at $\theta=0$ and at the value used before to 
determine $c$. In Fig.\ \ref{figure3} we show 
how well this works in the $^{208}$Pb example.  Plotted
as circles are the ratios $\lb q | H R_\theta | q \rb /\lb q | R_\theta| q \rb$
calculated from the many-particle wave functions.  The fits by the 
improved GOA
and the topGOA are shown as curves.  Again the topGOA is superior at
the moderate deformation of this example. 

We now carry out the integrals to get the angular momentum projected
energies
\be
E_{qq',J=0} 
= 2 \, I^{-1}_0 (q,q') \int_0^1 d \cos (\theta) \, \lb q| H R_\theta|q'\rb
.
\ee   
We report here the diagonal correlation energies $\Delta E_{q,proj}$
defined by the difference in the projected energy and the energy of
the intrinsic configuration,
$$
\Delta E_{q,proj} = E_{qq,J=0} - \lb q | H | q \rb
.
$$
Later, we shall present the energy difference with respect to 
the minimum energy of the unprojected configurations,
\be
\label{Ecorr}
\Delta E_{corr} 
= \min_q E_{qq,J=0} - \min_q \lb q | H | q \rb.
\ee
This correlation energy is more useful to assess the effect of
the projection on the calculation of the ground state binding
energy.

\begin{figure}[t]
\centerline{\epsfig{figure=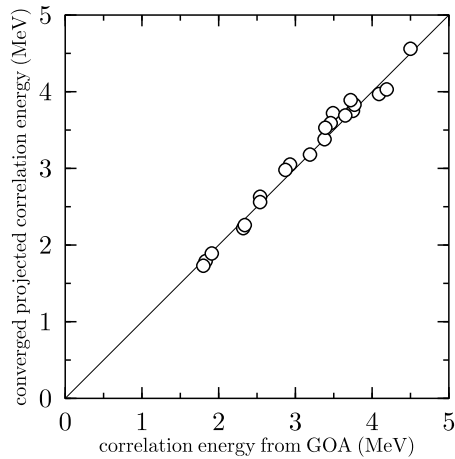}}
\caption{\label{figure4}
Comparison of explicit and GOA calculations of the projection,
for the nuclei \sn120 and \sm156 and a number of deformations.
}
\end{figure}

For the case presented in Figs.\ \ref{figure2} and \ref{figure3}, 
the energy gained by projection
is 5.8 MeV for the accurate integration, and essentially the same
value for the GOA.  In this case, both the topGOA and the naive
GOA give values within 0.1 MeV of each other. 
In Fig.\ \ref{figure4} we show the comparison of the 
(numerically converged) 12-point
projection with the GOA for a number of other nuclei and deformations.

The correlation energy associated with the angular momentum projection
has a range of about 2 to 4 MeV, with the larger
values occurring for well-deformed configurations.  The topGOA
tracks these values closely, with a typical error of the order
of 100-200 keV.  

\begin{figure}[t]
\centerline{\epsfig{figure=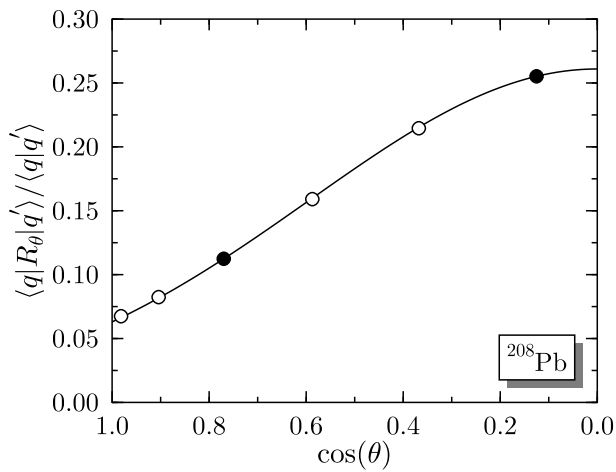}}
\caption{\label{figure5}
Same as Fig.\ \ref{figure2}, for the off-diagonal overlap $\lb q | R_\theta
| q' \rb$ with $q=500$ fm$^2$ and $q'=-500$ fm$^2$, which corresponds
to $|\beta_2| = 0.063$.
}
\end{figure}

In the above, we only examined overlaps and matrix elements between
configurations with the same sign of deformation.  These quantities
are of course also needed for matrix elements between oblate
and prolate configurations.  However, due to the choice of the 
same symmetry axis for oblate and prolate configurations, the 
overlap function has a rather different appearance.  An example 
is shown in Fig.\ \ref{figure5}, the overlap between
the configurations $q=500$ and $-500$ fm$^2$ of the nucleus \pb208.  Here
we see that the overlap peaks at 90$^\circ$.  This is not surprising.
The maximum density overlap 
is obtained by aligning the long axis of the prolate ellipsoid to one of
the longer axes of the oblate ellipsoid.  This requires rotating one of
the ellipsoids by 90$^\circ$.  We may continue to use the functional
form of the topGOA in this case, but the first point should be
taken where the overlap is maximum.  Thus, two rotations are
needed, a first one by 90$^\circ$ for the maximum, 
which is equivalent to a permutation of axes,
and at some intermediate
angle to get the width of the Gaussian.  The curve in Fig.\ \ref{figure5} 
shows the fit obtained using the two points shown by filled circles.

\begin{figure}[t]
\centerline{\epsfig{figure=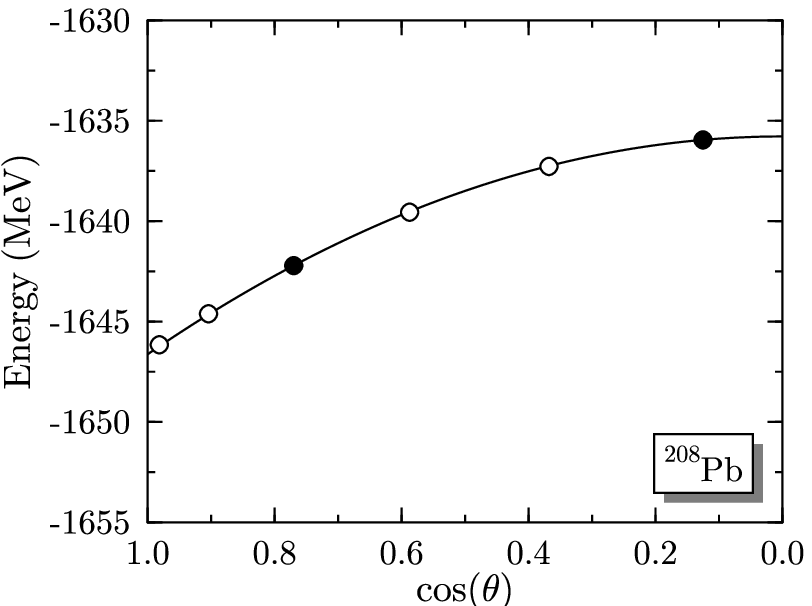}}
\caption{\label{figure6}
Same as Fig.\ \ref{figure3}, for the off-diagonal Hamiltonian matrix element 
$\lb q | H R_\theta
| q' \rb$ with $q=500$ fm$^2$ and $q'=-500$ fm$^2$.
}
\end{figure}

We determine the two parameters of the Hamiltonian matrix 
elements, eq.\ (\ref{hgoa}), using the same two angles. The fit is 
compared to the numerically calculated values in Fig.\ \ref{figure6}.  
We can define a correlation energy for the off-diagonal
matrix elements similar to the definition for diagonal elements,
$$
\Delta E_{q,-q';proj} = E_{q,-q';J=0} - {\lb q| H R_{\pi/2}
| -q' \rb \over\lb q | R_{\pi/2} | -q' \rb}.
$$
For the case presented in Figs.\ \ref{figure5} and \ref{figure6}, 
the value calculated with the fine angular mesh is 2.50 MeV, 
to be compared with 2.49 MeV estimated by the topGOA. For 
crossover matrix elements such as this one, one should not 
expand about $\theta=0$ using a nontopological GOA.

It is possible to use one of the other GOA's if one makes 
a rotation of one of the intrinsic configurations by $90^o$
to align the longer axes. Calling the rotated wave function 
$|q,\pi/2\rb \equiv R_{\pi/2}|q\rb$, the overlap integral becomes 
\be
\label{I90}
I_0(q,q') 
= 2 \int_{0}^1 d\sin (\theta) \lb q,\pi/2| R_\theta |q'\rb.
\ee
In this representation, the improved GOA gives a value for the 
overlap which differs by less than 1\% from the value obtained
by 12-point Gauss-Legendre integration.
Note that topGOA gives identical values 
with either eq.\ (\ref{exactintegral}) or (\ref{I90}).

\section{Mixing deformations}

In the previous section, we have presented a numerical approximation
which allows one to calculate the matrix element projected on 
$J=0$ between any two intrinsic configurations $(q,q')$. If the 
total number of configurations is $N$, one needs
$N(N+1)/2$ elements per matrix. In this section,
we introduce an approximation scheme which requires only that
diagonal and nearest-neighbor off-diagonal matrix elements be
calculated with the full many-particle wave functions.  This
reduces the number of needed elements to $2N-1$, giving  
considerable savings in computational cost.

The diagonalization of the discretized Hill-Wheeler equations 
is equivalent to the variational theory 
based on a wave function of the form
$$
|\Psi\rb = \sum_i c_i \, |q_i,0\rb
,
$$ 
where the $c_i$ are variational parameters and
$|q_i,0\rb$ are the $J$-projected intrinsic configurations.  The total
energy is thus
$$
E_{HW} = \min_{c_i} {\lb \Psi |H| \Psi\rb\over\lb\Psi| \Psi \rb}
$$
and the energy gained by configuration mixing is
\be
\label{E_HW}
\Delta E_{HW} = E_{HW} -\min_q E_{qq,J=0}.
\ee
As mentioned earlier,
the configurational basis only includes axially symmetric deformations.
For a given sign of the deformation (prolate
or oblate), the intrinsic states can be ordered with respect to their
quadrupole moments.  The diagonal and nearest-neighbor off-diagonal
elements must be calculated from the full mean-field wave functions, but
the remaining matrix elements can be estimated using the GOA as follows.
First consider the overlaps.  The 
number and angular-momentum projected wave functions are renormalized to unity
in the formulas below, i.e.\ we take $I_0(q_i,q_i) = 1$.
The overlaps between neighboring
states $I_0(q_i,q_{i+1})$ are then used 
to define a variable $x$ \cite{bo90},
\be
\label{xmap}
x_i=x_{i-1} + \sqrt{-2 \log (|I_0(q_i,q_{i+1})|)}.
\ee
This plays the role of a coordinate for the GOA.  Accordingly,
the overlaps between more distant states are estimated as
\be
\label{eq7}
I_0(q_i,q_j) \approx \exp( -(x_i-x_j)^2/2).
\ee

\begin{figure}[t]
\centerline{\epsfig{figure=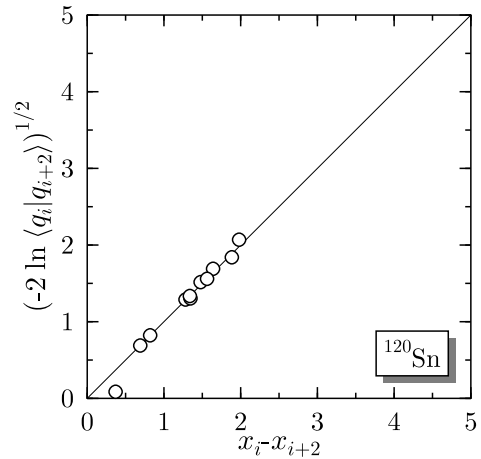}}
\caption{\label{figure7}
Overlaps of angular-momentum and particle-number projected states
from second-nearest neighbor configurations of \sn120.  The 
ordinate and abscissa indicate the actual and the GOA-estimated
overlaps, respectively.
}
\end{figure}

Let us see how well this works for the configuration set 
used for $^{120}$Sn.  The set contains 13 states, with
deformations covering the range from strongly oblate 
($q$=-1000 fm$^2$) to strongly prolate ($q$=1300 fm$^2$). 
A typical separation between neighboring states is $x_{i+1}-x_i\sim 0.8$.  
Fig.\ \ref{figure7} shows the overlaps between the second-nearest
neighbors, plotted as an equivalent separation
$y_i^{(2)}=\sqrt{-2 \log (|I_0(q_{-1},q_{i+1})|)}$.
The $x$-axis gives the separation as determined from the assigned
$x$ values from eq.\ (\ref{xmap}), $\Delta x = x_{i+1}-x_{i-1}$.
The accuracy of this GOA can 
be judged by the deviation of the points from the diagonal line.
We see that eq.\ (\ref{eq7}) gives satisfactory results except 
for one point. The bad point corresponds
to the oblate-to-prolate matrix element $I_0(200,-200)$;
all other points correspond to transitions between states have
the same sign of deformation.  
The prolate-oblate matrix element has a value very close to unity, 
$I_0(200,-200)=0.996$ and $y^{(2)}=0.08$, 
despite the fact that the overlaps with the spherical state that
separates them is smaller ($\Delta x = 0.17$ and 0.19, for the
prolate and oblate states, respectively).

Unexpectedly large overlaps between prolate
and oblate projected states were also found in the previously
mentioned study of the interacting boson model \cite{ha03}.
In fact it is easy to understand this behavior.  For small
deformations, the intrinsic configuration may be expanded
as $ |q\rb \approx |0\rb + \alpha Q^\dagger |0\rb + \alpha^2
Q^\dagger Q^\dagger |0\rb/2 + \ldots$ where $Q$ is a particle-hole operator
transforming as $J=2$, $M=0$ under rotations and $\alpha$ is a
deformation parameter.  On projection, the wave function becomes
$ P_{J=0} |q\rb \approx |0\rb + \alpha^2/2\sqrt{5} (Q^\dagger Q^\dagger)^0
|0\rb + \ldots$.  This wave function does not depend on the sign of $\alpha$.
Thus, the same state can be obtained by projecting a prolate or
an oblate intrinsic state.

\begin{figure}
\centerline{\epsfig{file=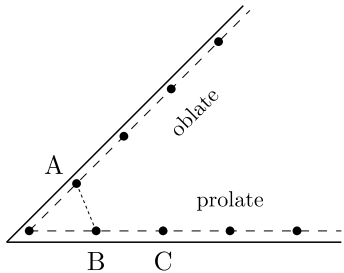}}
\caption{\label{figure8}
Linear metrics for GOA mixing prolate and
oblate shapes. The naive metric is defined by the path passing
through the spherical point, shown by the dashed line in the
figure.  An improved metric bypasses the spherical point via
the path AB.
}
\label{naive}  
\end{figure}

This behavior is also found for crossover matrix elements between more
highly deformed configurations, and it
prevents us from defining a linear metric $x$ to
estimate the crossover matrix elements. A possible procedure could be
to carefully select the sequence of deformations 
so that eq.\ (\ref{eq7}) is not applied to estimate high-overlap 
crossover matrix elements. This strategy is illustrated in 
Fig.\ \ref{figure8}. The dashed line shows the path for a naive 
scheme for assigning a distance between configurations.

The points marked $A$ and
$B$ would be separated by a distance $\Delta x$ given by the sum of
the distances between the origin and those two points.  But from
what we said before, the angular-momentum projected configurations
$A$ and $B$ would
actually have a very high overlap, larger than with the spherical 
configuration.  Modifying the path by taking
the short-cut $AB$ (dotted line), the overlap $I_0(q_A,q_B)$
will be determined exactly and the errors for more distant crossover
overlaps will be reduced.  This procedure offers an improvement, but
more distance crossover overlaps such as $AC$ are still underestimated.
This suggests that the two-dimensional geometry of 
Fig.\ \ref{figure8} might be used to define a better metric 
for the overlaps:  measure the distance along the lines for the 
prolate and oblate configurations separately as done before, but 
use the triangular geometry of Fig.\ \ref{figure8} to 
determine crossover overlaps.  We put the prolate configurations
along the horizontal axis, and write their position as a vector
$\vec x_i = (x_i,0)$. The oblate configurations are put on a
line making an angle $\alpha$ with respect to the horizontal axis,
located at positions $\vec x_i = x_i (\cos\alpha,\sin\alpha)$.
The distance between 
configurations $q_i,q_j$ is taken to be $y_{ij} = |\vec x_i - \vec x_j|$, 
which is to be applied in general without distinguishing 
prolate and oblate.  We will call this the \emph{triangular GOA}, 
and will apply it below.  In general, the angle $\angle A0B$
of the triangle depends on the distances, becoming more acute
the closer $A$ and $B$ are to the spherical point.  For the
configuration sets used here, the spacings are such that
the angles range from $30^\circ$ to $60^\circ$.  In the application
below, we assume a fixed angle of $45^\circ$.

We now turn to the estimation of the Hamiltonian matrix element.
The first step for a GOA estimate is to express
the matrix element as a product
of the overlap and a smoothly varying factor 
$h(q_i,q_j)$,
$$
\lb q_i | H | q_j \rb = h(q_i,q_j) I_0(q_i,q_j).
$$
The factor $h$ is expanded as a power series in 
$|\vec x_i-\vec x_j|$,
\be
\label{h0,2}
h(q_i,q_j) =    h_0(\bar x) - h_2(\bar x)|\vec x_i-\vec x_j|^2,
\ee
where $\bar x = (x_i+x_j)/2$.  This equation will be used to define
$h_2$ between nearest-neighbor configurations $q_i,q_{i+1}$, taking the 
computed Hamiltonian matrix element as input.
To assign nearest neighbors, we drop the spherical point and use the 
dotted path shown in Fig.\ \ref{figure8}. The definition of 
$h_2$ on these links is thus 
\begin{eqnarray}
h_2(i,i+1)
& = & \frac{1}{|\vec x_i -\vec x_{i+1}|^2} \bigg(
      \frac{ h(q_i,q_{i})+ h(q_{i+1},q_{i+1})}{2}   
      \nonumber \\
&   & - h(q_i,q_{i+1}) \bigg)
.
\nonumber
\end{eqnarray}
We average the $h_2$ on the links between two states to get an estimated
$\bar h_2$ to 
use in their matrix element.  Then the matrix element of the Hamiltonian
on the $J-$projected states is estimated as
\be
\label{h_estimated}
h(q_i,q_j) 
= \frac{1}{2} \big[ h(q_i,q_i)+h(q_j,q_j) \big] 
  + \bar h_2 \, |\vec x_i -\vec x_j|^2.
\ee

We now have complete matrices both for the overlap and the Hamiltonian
and it is diagonalized exactly the same way as in the full theory. 
This is usually done by first diagonalizing the matrix of overlaps $I_0$ and constructing
its square root.  Then the matrix
$$
I_0^{-1/2} H I_0^{-1/2} 
$$
is then diagonalized, and its eigenvalues give the energies of
the GCM theory. We remind the reader that
the method has numerical instabilities if the configurations are too
close. This is circumvented by truncating the basis, removing
states whose norm is small (typically $\le 0.01$). In such situations, 
there is a sensitivity to elements of the norm matrix that are quite
removed from the main diagonal.  It is for this reason that the
GOA needs to be rather accurate. As an example, the configuration set
used to represent \sn120, three of the thirteen eigenstates of the norm
matrix had eigenvalues less that $0.01$ and were discarded.

\begin{figure}[b]
\centerline{\epsfig{figure=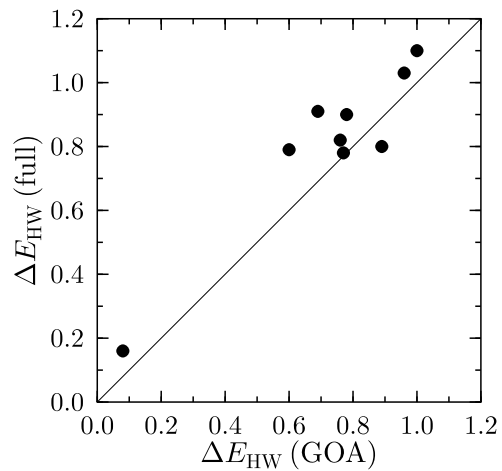}}
\caption{\label{fig9}
Comparison of configurational correlation energies $\Delta E_{HW}$, 
eq.\ (\ref{E_HW}), computed with the triangular GOA to the numerically 
accurate values.  
}
\end{figure}

We now compute the correlation energy for a sample of nuclei using
the triangular GOA and compare with the results of the fully calculated
matrices. The resulting configurational correlation energy
eq.\ (\ref{E_HW})
is displayed in Fig.\ \ref{fig9}. The horizontal axis shows the value
in the GOA using eqs.\ (\ref{eq7}) and (\ref{h_estimated}), 
and the vertical axis shows the accurate result of the full calculation.

The errors using the estimated matrix elements are all within the
targeted accuracy of $\pm0.2$ MeV.  
We conclude that the procedure has
sufficient accuracy to be used in a first survey of the nuclear mass
table.

For numerical purposes in general, it is also important to know how 
close together the configurations need to be to get reliable energies.  
Ref.\ \cite{bo90} indicates
that $\delta x= 1.6 $ (overlap of 30\%) is adequate and $\delta x = 1.1$
(overlap of 55\%) is safe, taking a criterium of 0.2 MeV for energy
convergence. This seems to be valid for the $J-$projected configurations
as well. For example, for the nucleus \sn120,  we can thin the 
configuration set taking out states whose
overlaps with neighbors exceeds 55\%.  Here the thinned space has only
4 configurations, compared with 13 in the original set.  The configurational
correlation energy is nearly unchanged by the thinning and is within
the desired error bound.  Clearly, more study
is needed to find an optimum spacing for a global prescription.

\section{Examples of correlation energies}

In this section we survey the trends of the correlation energy 
taking the results for a number of nuclei. In this sampling we
have chosen nuclei to represent different kinds of structure in the
mean field approximation, including closed-shell magic (\pb208),
strongly deformed (\sm156), and semi-magic with strong pairing (\sn120).  
In addition, we consider two isotope chains, krypton and lead, which
exhibit coexisting oblate and prolate deformations. The results
are presented in Table \ref{tab:corrGCM}. The first three columns of 
energies show the numerically converged correlation energies obtained with 
the $N$, $Z$, and $J$ projections and Hill-Wheeler representation as 
in refs.\ \cite{va00,du03,BFH03,be03}.

The first column gives $\Delta E_{corr}$, the energy gain over that of the 
lowest intrinsic configuration associated with angular momentum projection,
defined in eq.\ (\ref{Ecorr}).  We should note that we only  use the
given configurations to determine the energy minimum for the projected
states,  $ \min_q E_{qq,J=0}$, which could give some
error in assigning the energy gain. There will be a compensating
error in the value of the configurational correlation energy, with
the total independent of $ \min_q E_{qq,J=0} $.
In the table, the nuclei for which the deformation $q$ is
significantly different for the intrinsic and the projected configurations 
are indicated with asterisks.  An example where the deformation is
virtually identical before and after projection is \sm156.  Its  
correlation energy is 3.0 MeV, which is rather typical for deformed
nuclei. What is surprising
to us is that the other nuclei have rotational correlation energies of 
the same order of magnitude.  The semi-magic ``spherical" nucleus 
\sn120 is a case in point.  At the mean-field or intrinsic level, the ground
state is indeed spherical, but the energy after angular momentum
projection is lower for the configuration with $q=400$ fm$^2$.  
Qualitatively this behavior is to be
expected, because the variation-after-projection is a way to
introduce correlations from the attractive parts of the interaction
that only have perturbative character.  The surprising
finding is the magnitude, which is only 10\% lower in \sn120 than
the value for a fully deformed nucleus.  In fact all the
nuclei except the doubly-magic \pb208 have correlation energies
in the same range.  

\begin{table}
\caption{\label{tab:corrGCM}
Correlation energies in the GCM associated with quadrupole
deformations.  Energies are given in MeV.}
\begin{tabular} {cc|cccc}
Z & A & $\Delta E_{corr}$  &  $\Delta E_{HW}$  & total & GOA\\
\tableline
36 &  74 & 2.9     & 0.8 & 3.7 & 3.7 \\
36 &  78 & 2.4$^*$ & 0.9 & 3.3 & 3.2 \\
36 &  82 & 3.0     & 0.8 & 3.8 & 3.5 \\
36 &  86 & 2.8     & 0.4 & 3.3 & 3.0 \\
50 & 120 & 2.8$^*$ & 0.8 & 3.6 & 3.3 \\ 
62 & 156 & 3.0     & 0.8 & 3.8 & 3.8 \\
82 & 186 & 2.7     & 0.9 & 3.6 & 3.3 \\
82 & 190 & 2.7     & 1.0 & 3.8 & 3.7 \\
82 & 194 & 2.8     & 1.1 & 3.9 & 3.7 \\
82 & 208 & 1.5$^*$ & 0.2 & 1.7 & 1.6 \\
\end{tabular}
\end{table}

The next column of the table shows
additional energy gain obtained by solving the Hill-Wheeler equation
to mix configurations, $\Delta E_{HW}$ in eq.\ (\ref{E_HW}).  
Except for the doubly-magic \pb208, these numbers are remarkably 
constant over the nuclei considered, ranging in value from 0.8 to
1.1 MeV. The nucleus \pb208 behaves in a similar way as the
model discussed in ref.\ \cite{ha03}.  That model permitted a
significant energy by projection, but the optimal projected state
had a very high overlap with the true eigenstate, and there was
no further energy gain by mixing configurations.

The third column shows the total correlation energy, 
$\Delta E_{corr} + \Delta E_{HW}$. The range of variation
is 0.6 MeV, except for \pb208.  

The last column in Table \ref{tab:corrGCM} shows the total correlation 
energy computed with the approximations in the treatment of the
Hill-Wheeler equation described in the last section.  The
results are very close to that of the full treatment, with
the r.m.s.\ error having a value of 0.15 MeV.  We now come
to the conclusions that we draw from these results.
  
\section{Conclusions}

Although the primary purpose of this study was numerical and we
only examined a small sample of nuclei, our
findings suggest the possible physics that may emerge from a systematic,
global study of the quadrupolar correlation energy.  One might
naively expect that the energy of projection would be large for 
deformed nuclei but not for others. In fact we found that this energy 
is rather large
except for the doubly magic nucleus that we considered. The additional 
energy associated with vibrations is small, less than one MeV, and
not very different from one nucleus to another.  For the sum
we see fluctuations of the order a few hundred keV.  

In the global theories of nuclear masses, one treats the mean-field
effects in detail but up to now ignoring or approximating in
a phenomenological way fluctuating parts of the correlation 
energy.  The relative constancy of our calculated correlation
energy suggests that these fluctuations are indeed small, permitting
a mean-field treatment to be quite successful.  On the other hand,
the fluctuation effects we have found are large enough to be
potentially useful.
In the present mass theories there are r.m.s.\ deviations of the
order of 700 keV.  For the nine nuclei in Table \ref{tab:corrGCM}, 
the r.m.s.\ fluctuation in the total correlation energy is 
650 keV. This gives grounds for the hope that 
including the quadrupolar correlation energy in calculating
nuclear masses might significantly improve the agreement with
the experimental values and produce a theory with better predictive power.

On the numerical side, the GOA for angular momentum projection reduces
the computational cost by almost an order of magnitude.  A similar
saving is realized by the triangular GOA for the Hill-Wheeler matrices, 
which only requires computing diagonal and next-to-diagonal elements. This
brings the total cost down the same level as the HF+BCS computations
of the GCM configurations, and makes feasible a systematic determination
of correlation energies for the nuclear mass table.

We thus have the intention of continuing this project to compute
the quadrupolar correlation energies of all even-even nuclei.  For
a global calculation we would need to use an energy functional that
is applicable across the chart of nuclides.  We have been using
the SLy6 Skyrme parameterization, but so far the pairing parameters 
have been different for
light and heavy nuclei.  Ultimately, a new fit of the mean field
energy functional will be required, but for a first survey of 
the correlation energy it should be sufficient to take the 
same functional, but with a fixed pairing interaction.

\section*{Acknowlegments}
We wish to acknowledge helpful discussions with H.\ Flocard,
P.-G.\ Reinhard, and K.\ Hagino.
GFB acknowledges support by the US Department of Energy under 
Grant DE-FG-06-90ER40561, the Guggenheim Foundation,
and the Institut de Physique Nucl\'eaire d'Orsay.
PHH thanks the Institute for Nuclear
Theory in Seattle, where this work was initiated. This work
has been partly supported by the \mbox{PAI-P5-07} of the
Belgian Office for Scientific Policy. 
MB acknowledges support through a European Community Marie 
Curie Fellowship.

\section*{Appendix: Summary of computational costs}

Our aim is to reduce the computing time needed to calculate
correlation energies at a level comparable to that of mean-field calculations.
Let us estimate the cost of both and to determine how they scale
with the number of nucleons. The parameters which govern the computing 
time are the number of active orbitals $N_{act}$ and the
dimension of the vector representing an orbital $N_r$. The mesh size 
is always fixed to 0.8~fm, a value sufficient to have a satisfactory
accuracy on energies. The number of mesh points must be 
large enough to guarantee that the wave functions can be set
equal to zero at the edge of the box so that derivatives and 
the Coulomb boundary conditions can be calculated accurately.
The mean-field equations are solved
by the imaginary time method and one can limit the number of orbitals 
that have to be computed to the orbitals close to the Fermi level. 
In practical applications, to be on the safe side, $N_{act}$ 
and $N_r$ are chosen larger than required.

In Hartree-Fock
theory $N_{act}$ is equal to $N_{occ}$, the number of occupied orbitals.
Due to time-reversal symmetry, the orbitals are two-fold degenerate 
and the number of orbitals is half the number of nucleons.
When pairing correlations are taken into account, either at the BCS 
or at the Bogoliubov level,  $N_{act}$ is larger. In practical applications,  
to be sure that enough
HF wave functions are computed, $N_{occ}$ is taken around $2A/3$.

Calculations are performed in a 3-dimensional coordinate grid
representation. To calculate the nuclear ground state and to be able
to easily rotate the orbital wave functions on the mesh, we consider
only cubic meshes. 
In practice, the actual dimension of the grid
can be reduced by symmetry considerations.
The reflection symmetry of the mean-field Hamiltonian allows one
to consider only the points in an octant. We will thus define $N_x$
to be the number of points along the positive $x$-axis. However,
there is a price to pay for using this symmetry: 
the wave functions must be complex.  Also, the spin
degree of freedom doubles the size of the vector.  In the
end the dimension of a real array representing an orbital is
$$
N_r =  4 N_x^3.
$$ 
The number of points scale roughly as $3A^{1/3}$, although the number used
in practical applications is larger for light nuclei. This dependence
on $A$ ensures that the wave functions can be safely put to zero at
the edge of the box.
   
The mean-field calculation is done iteratively, determining in each cycle
the total energy
of the nucleus, the HF Hamiltonian and its action on the individual
wave functions. These computations need to be repeated a number of time
equal to the total number of iterations required for a full convergence,
typically 200 to 400 times. It has also to be done for each
quadrupole moment which will enter the GCM calculation.
The GCM requires calculating the overlaps and the Hamiltonian
matrix elements between wave functions with different orientations 
and deformations. Let us first 
evaluate the computational effort that is required for a single 
set of Euler angles and a single deformation. This effort is similar
to that required to calculate  the energy in the mean-field
calculation.

For the calculation of a matrix element of the GCM
Hamiltonian kernel or the total HF energy 
the dominant terms are those with
derivatives of the wave functions. This has a computational cost given by
\be
\label{kinetic}
 6 N_{act} N_r N_x, 
\ee
following the method of ref.\ \cite{ba86}.  A prefactor 3 accounts for
the three Cartesian directions along which derivatives have to
be calculated. The prefactor 2
counts separately the multiplication of the initial vector and
the addition to the final vector.  Several terms
in the Hamiltonian matrix element involve derivations, e.g. those associated
with the spin-orbit field, and the total computational effort is
an order of magnitude larger than given by the expression Eq. 
(\ref{kinetic}). More terms appear in the GCM calculation due to
the presence of vector densities which are equal to zero for diagonal
matrix elements.

The cost related to particle-number 
projection is not easy to evaluate, since this projection is intimately mixed
with the other part of the calculation. We use the method of ref. \cite{HBD93}
which requires one to multiply the $v$ factor of the Bogoliubov
transformation by a phase. The calculation of the derivatives of the
wave functions is not affected by this projection. However, all densities
become complex.

The mean-field calculation requires one to calculate 
also the action of the HF Hamiltonian on all mean-field wave functions.   
Once more, the most time-consuming part concerns terms including
derivatives. They are grouped in terms with first and second
order derivatives acting on the four components of the wave functions,
with 24 terms in total.

In the GCM case, another time-consuming task is rotating the many-particle
wave function, an operation that is needed for angular momentum
projection.  This is done by interpolating the single-particle 
wave functions in the $x-z$ plane to the points corresponding
to some rotation $\theta$ about the $y$-axis, 
$$
R_\theta:\,\,\, (x,z) 
= (x'\rightarrow x\cos\theta +z\sin\theta ,
   z'\rightarrow -x\sin\theta +z\cos\theta )
.
$$
The needed accuracy is
achieved by using all the points in a half plane to interpolate
in that plane. The computational cost here is
$$
4 N_{act} N_r N_x^2
$$
where the factor 4 includes a factor 2 for expanding from a 
quarter-plane slice of the octant to the half plane.

To take overlaps, one needs to calculate a determinant.  The
determinant is of order $N_{act}$ and setting up its matrix
requires  
$$
2 N_{act}^2 N_r
$$
operations.  Since $N_r$ is usually less than $N_{act}$, this
is usually the most time consuming task.  However, the simultaneous 
projection on
particle number and angular momentum requires also additional calculations
to follow the phase of the overlap kernel. Test calculations indicate that
the particle number projection approximately doubles the computing of the
determination of the GCM  kernels.

For our calculations, $N_x$ is of the order of 14 (Kr) to 16 (Pb).  
Thus the vector size is about 
$N_r \approx 4 * 15^3 = 13,500. $ The number of orbitals for Pb is around
$N_{act} = 140$. With these numbers, the interpolation of
the wave functions has about the same cost as setting up the determinants.

Putting all these numbers together with realistic values, one can
estimate that the determination of a single point for the GCM+projection 
calculation takes a factor 10 longer than a single mean-field iteration.

The  computing time is also determined by the number of times these basic 
calculations have to be done. It is the number of iterations (typically 300) 
times the number of quadrupole moments that have to be considered
(typically 15) for the mean-field part. For the GCM part, it is the
number of active Euler angles (typically 12) times the number of 
matrix elements (typically $15^2$). With all these numbers, the GCM 
calculation is an order of magnitude longer than the mean-field calculation. 
Reducing the number of Euler angles to 1 or 2 and the number 
of matrix elements to be determined exactly to 2 times the number 
of discretized quadrupole moments reduce this part of the 
calculation to a time similar to the mean-field calculation.

\end{document}